\documentclass{frontiersFPHY} % for Physics and Applied Mathematics and Statistics articles

\usepackage{url,hyperref,lineno,microtype}
\usepackage[onehalfspacing]{setspace}
\usepackage{dirtytalk}

\def\keyFont{\fontsize{8}{11}\helveticabold }
\def\firstAuthorLast{Cihon {et~al.}} %use et al only if is more than 1 author
\def\Authors{Peter Cihon\,$^{1}$ and Taha Yasseri\,$^{2,*}$}
\def\Address{$^{1}$Williams College, Williamstown, MA, USA
$^{2}$Oxford Internet Institute, University of Oxford, Oxford, UK}

\def\corrAuthor{Taha Yasseri}
\def\corrAddress{Oxford Internet Institute, University of Oxford, 1 St Giles, Oxford, OX13JS, UK}
\def\corrEmail{taha.yasseri@oii.ox.ac.uk}

\begin{document}
\onecolumn
\firstpage{1}

\title[Twitter studies on political action]{A biased review of biases in Twitter studies on political collective action} 

\author[\firstAuthorLast ]{\Authors} %This field will be automatically populated
\address{} %This field will be automatically populated
\correspondance{} %This field will be automatically populated

\extraAuth{}% If there are more than 1 corresponding author, comment this line and uncomment the next one.
%\extraAuth{corresponding Author2 \\ Laboratory X2, Institute X2, Department X2, Organization X2, Street X2, City X2 , State XX2 (only USA, Canada and Australia), Zip Code2, X2 Country X2, email2@uni2.edu}

\maketitle

%%%%%%%%%%%%%%%%%%%%%%%%%%%%%%%%%%%%%%%%%%%%%%%%%%%%%%%%%%%%%%%%%%%%%%%%%%%%%%%%%%%%%%%%%%%%%%%%%%%%%%%%%%%%%%%%%%%%%%%%%%%%%%%%%%%%%%%%%%%%%%%%%%%%%%%%%%%%%%%%%%%%%%%%%%%%%%%%%%%%%%%%%%%%%%%%%%%%%%%%%%%%%%%%%%%%%%%%%%%%%%%%%%%%%%%
%%% The sections below are for reference only.
%%%
%%% For Original Research Articles, Clinical Trial Articles, and Technology Reports the section headings should be those appropriate for your field and the research itself. It is recommended to organize your manuscript in the
%%% following sections or their equivalents for your field:
%%% Abstract, Introduction, Material and Methods, Results, and Discussion.
%%% Please note that the Material and Methods section can be placed in any of the following ways: before Results, before Discussion or after Discussion.
%%%
%%%For information about Clinical Trial Registration, please go to http://www.frontiersin.org/about/AuthorGuidelines#ClinicalTrialRegistration
%%%
%%% For Clinical Case Studies the following sections are mandatory: Abstract, Introduction, Background, Discussion, and Concluding Remarks.
%%%
%%% For all other article types there are no mandatory sections.
%%%%%%%%%%%%%%%%%%%%%%%%%%%%%%%%%%%%%%%%%%%%%%%%%%%%%%%%%%%%%%%%%%%%%%%%%%%%%%%%%%%%%%%%%%%%%%%%%%%%%%%%%%%%%%%%%%%%%%%%%%%%%%%%%%%%%%%%%%%%%%%%%%%%%%%%%%%%%%%%%%%%%%%%%%%%%%%%%%%%%%%%%%%%%%%%%%%%%%%%%%%%%%%%%%%%%%%%%%%%%%%%%%%%%%%

\begin{abstract}

%%% Leave the Abstract empty if your article falls under any of the following categories: Editorial Book Review, Commentary, Field Grand Challenge, Opinion or specialty Grand Challenge.
\section{} 
In recent years researchers have gravitated to social media platforms, especially Twitter, as fertile ground for empirical analysis of social phenomena. Social media provides researchers access to trace data of interactions and discourse that once went unrecorded in the offline world. Researchers have sought to use these data to explain social phenomena both particular to social media and applicable to the broader social world. This paper offers a minireview of Twitter-based research on political crowd behavior. This literature offers insight into particular social phenomena on Twitter, but often fails to use standardized methods that permit interpretation beyond individual studies. Moreover, the literature fails to ground methodologies and results in social or political theory, divorcing empirical research from the theory needed to interpret it. Rather, papers focus primarily on methodological innovations for social media analyses, but these too often fail to sufficiently demonstrate the validity of such methodologies. This minireview considers a small number of selected papers; we analyze their (often lack of) theoretical approaches, review their methodological innovations, and offer suggestions as to the relevance of their results for political scientists and sociologists. 

\tiny
 \keyFont{ \section{Keywords:} social media, twitter, mobilization, campaign, collective action, bias, theory} %All article types: you may provide up to 8 keywords; at least 5 are mandatory.
\end{abstract}

\section{Introduction}

% For Original Research Articles, Clinical Trial Articles, and Technology Reports the introduction should be succinct, with no subheadings.
%
% For Clinical Case Studies the Introduction should include symptoms at presentation, physical exams and lab results.
%
Since its founding in 2006, Twitter has become an important platform for news, politics, culture, and more across the globe \cite{weller2014}. Twitter, like other social media platforms, empowers new forms of social organization that were once impossible. Margetts et al. 2016 discuss changing conceptions of membership and organization on social media \cite{margetts2015}; Twitter communities and conversations need not be bounded by geography, propinquity, or social hierarchy. As a result, social and political movements have taken to the site as a means of organizing activity both online and offline. In facilitating these movements, Twitter simultaneously makes available a data trail never before seen in social research. Researchers have embraced these data to create an expanding body of literature on Twitter and social media writ large.

Yet, this body of literature is only unified in the source of its data; it remains fractured across many disciplines and fails to establish set procedures for drawing conclusions from these rich datasets. Indeed, metareviews of election prediction using Twitter have raised significant concerns of this literature's validity \cite{gayo2012,metaxas2012}. This minireview extends this critical discussion of Twitter literature to political action. We selected the reviewed literature in order to sample a variety of topics and methodologies. The sample purposively draws a geographic diversity of papers studying Twitter-based political action in Europe, the Middle East, and the United States. Yet, some gaps certainly remain, including the glaring absence of hashtag activism studies and terrorist propaganda activity, two topics important to political action on Twitter that warrant further study. Hence we acknowledge that our review is not inclusive in terms of coverage of all the relevant papers in the field. 

In reviewing the state of Twitter literature on political action, we seek to explain the role of computational social science (also called social data science) methodologies in augmenting political scientific and sociological understanding of these phenomena. Our minireview is structured as follows. We begin by examining the role of theory, and find that most often authors do not consider the expansive political and social theoretical literature in their analyses of online social phenomena. Instead, they provide case studies and methodological developments exclusively for Twitter research. We next examine the methodologies of these studies, and, drawing upon Ruths et al. \cite{ruths2014}, we find that many papers fail to support their choice of methodology within the greater literature. We then examine significant results and discuss implications for further Twitter studies of political action.

\section{Where is the Theory?} \label{sec:theory}

Social and political theory serves an important role in making sense of social research by fitting individual studies into larger theoretical frameworks. In this way, individual studies can intelligibly inform future research. Alternatively, data analysis without a coherent, defensible theoretical framework serves only to explain a single observation at one point in time. The papers reviewed here fall into three broad categories in their use of theory: no theory, theory-light, and theory-heavy. Papers fall into these categories irrespective of methodological or phenomenological focus.

Papers without theoretical grounding may cursorily cite but fail to engage theoretical texts. Beguerisse-Dias et al. \cite{beguerisse2014} examine communities and functional roles on Twitter during the UK riots of August 2011. To explain these phenomena, however, they cite no social theory. While the authors offer sophisticated methodical innovations for determining interest communities and individual roles in those communities, they do so without reference to a broader social science literature. Some other papers offer cursory theory in their discussions of Twitter data. Borge-Holthoefer et al. \cite{borge2015} investigate political polarization surrounding the events that precipitated Egyptian President Morsi's removal from power in 2013. Their analysis of changes in “loudness” of opposing factions, although quite enlightening, is not grounded in any theoretical model of political action. Instead, the authors proceed based on a number of platform-specific assumptions that do not readily permit results to be generalized beyond Twitter. The authors suggest their findings contribute to the study of bipolar societies yet do not develop a theoretical model for such applications. The authors do use social theory, however sparingly, in order to contextualize their results, but even here theoretical discussion is lacking. Conover et al.\cite{conover2012} study partisan communities and behavior on Twitter during the 2010 U.S. midterm elections. They similarly prioritize analytical innovations over theoretical models. The authors analyze behavior, communication, and connectivity between users, but do not seek to explain observed partisan differences. Their research yields statistically significant differences between follower and retweet networks of liberal and conservative communities, which begs the question of why they exist? Such explanations could benefit from examining elections literature to develop a general theoretical model of partisan sharing. Although the authors do briefly address the 2008 U.S. presidential election, it is only to contrast resulting phenomena, not to offer explanatory theories.

In contrast, Alvarez et al. \cite{alvarez2015} explain political action in the Spanish 15M movement using Durkheimian theory of collective identity and establish their work on firm basis in collective action literature. Yet, while the authors base their methodology in theory, their findings do not directly engage with that theory aside from ``quantifying'' it. A similar fate befalls sampled predictive studies, which draw on theory to produce empirical results, but often fail to engage those results with underlying theory. Weng et al. \cite{weng2013} develop a model that predicts viral memes using community structure, based on theoretical insight from contagion theory. The authors find that viral memes spread by simple contagion, in contrast to unsuccessful memes which spread via complex contagion; still, only the briefest theoretical discussion for this result is offered. Garcia-Herranz et al. \cite{garcia2014} develop a methodological innovation using individual Twitter users as sensors for contagious outbreaks based in the ``friendship paradox'' and contagion theory. This mechanism uses network topology as an effective predictor, but does not address the social phenomena that create and sustain that topology. Such methodological innovations provide researchers new analytical tools for observational analysis, but these tools remain of dubious explanatory value because they fail to ground methods in theory of the social world.

Twitter data present an opportunity not simply for analysis of social interactions on the platform but, if done well, these insights hold potential to contribute to new visions of the social world. Rigorous data science can generate new theory. Coppock et al. \cite{coppock2015} are particularly notable in this regard. The authors base their methodological innovation in Twitter mobilization inducement on an extensive theoretical literature review, which yields three opposing hypotheses. They assess the political theory of collective action as it applies to Twitter via these three hypotheses, and find that the Civic Voluntarism Model is most consistent with their results. Likewise, Gonzalez-Bailon et al. \cite{gonzalez2011}, in their study of protest recruitment dynamics in the Spanish 15M movement, offer both an extensive grounding in social theory and theory-engaging results. The authors' findings serve to clarify threshold models of political action and ``collective effervescence''.

As to the particular theories addressed, the above mentioned papers focus primarily on political action and network theories of diffusion and contagion. Important in such topics, but absent from all papers, is discussion of power or hierarchy. Twitter operates within numerous contexts, e.g., the offline influence of particular users and the online influence of those with numerous followers. Reconciling methodologies with theories of power promises to provide further insight into political action on Twitter. More broadly, a greater focus on theory is needed for Twitter analyses to provide externally valid insight into the social world, both online and off. 

\section{Divergence in Methods}

In developing analyses of Twitter data, researchers have not drawn on a coherent body of agreed-upon methodologies. Rather, methodological choices differ considerably from one paper to another. Ruths et al. \cite{ruths2014} offers a critique of many common social media analysis practices. Drawing from that work as well as our own insights, we examine many of the methodological choices made in our sample papers. We have delineated these choices into several overarching categories: data, filtering, networks and centrality, cascades and communities, experiments, and conjecture.

Before addressing the methodological choices outlined above, we first address several important findings from Ref.~\cite{ruths2014}. Today, academic research writ large---including social media work and much more---is insufficiently transparent. Academic journals publish only ``successful'' studies. Without publishing methodologies that failed to explain political action phenomena, how is one to weigh the probability that the supposed ``fit'' observed is not due to random chance? Even those papers which address the robustness of their analysis, often stop at a very shallow significance tests using p-value, which is argued to be a flawed practice \cite{vidgen2016,wasserstein2016}. Similarly, when new methodologies are created, as in \cite{weng2013,garcia2014,coppock2015}, they are justified vis-\`a-vis random baselines and not prior methods. New methods are useful, but are they better than existing tools? These opacity critiques are fundamental to the current state of Twitter scholarship. Researchers should be cognizant of these limitations when drawing conclusions from their work and should alter their methodologies to account for these limitations whenever possible.

\subsection{Data}

Twitter data ultimately comes from the Twitter platform. If scholars wish to make claims about the versatility of their methodologies and findings, they must justify their data-collection methods as representative of underlying populations ---on Twitter or elsewhere. This proves a problematic task. The Twitter API offers researchers an incredible array of tweet, user, and more data for analysis; yet, the API acts as a ``blackbox'' filter that may not yield representative data \cite{morstatter2013a,morstatter2013b}. For example, Weng et al. \cite{weng2013} ``randomly'' collect 10 percent of public tweets for one month from the API. Not only does the API preclude analysis as to the representativeness of the sample but it too prevents researchers from comparing studies over time, as the API sampling algorithm itself will change. Proprietary sampling methods only further exacerbate the opacity problem. In \cite{gonzalez2011}, the authors use a proprietary sampling method to generate their dataset of Spanish tweets from Spain. The authors of \cite{borge2015} do as well, using Twitter4J and TweetMogaz as data sources.

Other papers do not use a global sampling method, but obtain data in other ways. Beguerisse-Dias et al. \cite{beguerisse2014} use a list of ``influential Twitter users'' published in {\it The Guardian} as the starting point for data collection. Coppock et al. \cite{coppock2015} develop their experimental design in cooperation with the League of Conservation Voters, and use their Twitter followers as test subjects. Other papers, including \cite{conover2012} and \cite{alvarez2015} collect data by following particular hashtags and the users who tweeted them. Garcia-Herranz et al. \cite{garcia2014} collect Twitter data by snowball-sampling from one influential user, Paris Hilton, as well as all users mentioning trending topics. None of these sampling methods allows authors to make broad claims about the Twitter platform and political action in general. The method used in \cite{garcia2014} is particularly concerning, as it attempts to collect a large sample to sufficiently model a Twitter population, but the choice of method undermines this very goal.

A final complication of data in Twitter studies regards the publication of that data. Once data is collected and analyzed, it is rarely made available for others to replicate these studies ---the hallmark of good research. The problem here lies with Twitter itself; the terms of use preclude the republication of tweet contents that have been scraped from the site.\footnote{Twitter Terms of Service: \url{https://twitter.com/tos?lang=en}} 

\subsection{Filtering}

Following data collection, researchers often filter an intractable dataset into a manageable sample. Researchers often use filtering to select a coherent sample. Language and geography offer clear examples. Borge-Holthoefer et al. \cite{borge2015} limit their dataset to Arabic tweets about Egypt. Both Gonzalez-Bailon et al. \cite{gonzalez2011} and Alvarez et. al \cite{alvarez2015} limit their datasets to Spanish tweets from Spain. To do so, however, both papers use a proprietary filtering process from Cierzo Development Ltd. As addressed above, proprietary methodologies stymie research transparency and replication. 

Filtering can likewise facilitate a narrowing of research focus given a particular sample population. One common means of achieving a relevant dataset is to use hashtags as labels for tweets in which they appear. In \cite{gonzalez2011} the authors obtain a sample of protest-related tweets using a list of 70 hashtags affiliated with the Spanish 15M movement. Conover et al. \cite{conover2012} filter to a sample of political tweets using a list of political hashtags and, in an excellent technique, allow the list of hashtags to grow based on co-occurring hashtags. In \cite{borge2015} the authours go one step further, and query not only hashtags but complete tweet content. Arabic tweets were normalized for spelling and filtered by a series of Boolean queries with a set of 112 relevant keywords. 

Researchers, after filtering for a relevant sample and topic, may further filter for user attributes. Borge-Holthoefer et al. \cite{borge2015} restrict their sample to high activity users with more than ten tweets extant in the limited sample. Beguerisse-Dias et al. \cite{beguerisse2014} limit their dataset to central users, those in the giant component of tweet-connected users. Weng et al. \cite{weng2013} limits the data to only reciprocal relationships. Conover et al. \cite{conover2012} filter tweets with geo-tags. The authors use a self-reported location field as their data source, despite the fact that someone can put ``the moon'' or anything else as their location. Although the authors acknowledge the preliminary status of their analysis and its utility as an illustration of potential data-driven hypotheses, it left us unsatisfied with a lack of methodological rigor that should underlie even the most tentative of filtering claims.

Authors may choose to filter for no other reason than to obtain a manageable dataset. Such decisions need not be arbitrary. Garcia-Herranz et al. \cite{garcia2014} settle on a particular sample size for their analyses, seeking to balance statistical power and the need to keep test and control groups from overlapping in the network. The authors offer an effective defense of their decision, presenting brief analyses of other sample sizes as well. Coppock et al. \cite{coppock2015}, on the other hand, arbitrarily remove Twitter users with more than 5000 followers from their sample because, they argue, these users are ``more likely'' to be influential or organizations, and therefore differ from the rest of the sample. This decision to remove outliers and the arbitrariness of the choice of threshold introduces systematic biases in the results, fundamentally undermining their analyses.

These myriad filtering decisions often go insufficiently defended. Those who do defend filtering choices often do so without referencing past literature. Even sound filtering decisions, however, undermine the general claims researchers can make. This may be one reason most of the studies fail to contribute to social theory beyond their micro case studies.

\subsection{Networks and Centrality}

Twitter lends itself to fruitful network analyses---of both explicit interactions and other derivative relations. Conover et al. \cite{conover2012} use three network projections to analyze partisan political behavior during the 2010 U.S. midterm elections: one network sees users connected when mentioned together in a tweet, another where users are linked by retweeting behavior, and, third, the original explicit user follow-ship network. Weng et al. \cite{weng2013} also uses three networks---mention, retweet, and follow---to study meme virality. The authors conduct primary analysis on the follow network and use the other two as robustness tests. In studying protest recruitment to the 15M movement, Gonzalez-Bailon et al. \cite{gonzalez2011} make use of two networks, one symmetric (comprised of reciprocated following relationships) and one asymmetric to study protest recruitment to the 15M movement. The authors use these networks to determine the influence of broadcasting users. Still other authors use single, traditional follower networks in their analyses \cite{garcia2014,coppock2015}.

Network analyses are all the more powerful when they are combined, as in \cite{weng2013} and \cite{gonzalez2011}. In \cite{borge2015}, the authors offer another insight when they use network analyses over time with temporally evolving networks in response to events that preceded Egyptian President Morsi's removal from power. The authors recreate a sequence of networks that evolve over time. This method offers insight into how online activity responds to offline events in Egypt, and could be a powerful tool in many other contexts, helping to parse a key question of political action: how groups respond to events and evolve over time. The opposite, to assume a network remains static during a given period, precludes this insight and undermines social analysis. In \cite{gonzalez2011}, the authors exemplify this pitfall, as a network of protesters being recruited surely saw significant changes during their study's time period. Given the fast growing literature on temporal networks \cite{holme2012,holme2015}, more attention is required in analyzing the dynamics of networked political activities. 

Beyond decisions of network type and temporality, authors make important choices in projecting and using Twitter networks. \cite{weng2013} does not weight network edges based on number of tweets, and choses to limits the network projection to reciprocal relationships. Both decisions fundamentally affect results, and undermine its validity as representing activity on the Twitter platform. Others, including \cite{gonzalez2011} account for asymmetry in their network projections. 

In doing network analysis, many researchers use centrality scores as a means to find the most influential users. Researchers have developed a number of different definitions and algorithms for centrality \cite{newman2010}. The choice of a specific approach, however, depends on the particular context and research questions. Often times this choice is not well justified in the given context of online political mobilizations. Among the papers considered here, k-core centrality \cite{wasserman1994} is the most common choice \cite{conover2012,gonzalez2011,alvarez2015}. While k-core centrality is a very useful tool to find the backbone of the network, it neglects social brokers, or the nodes with high betweenness centrality, ---relevant features in their own right when studying social behavior \cite{gonzalez2013}.

\subsection{Cascades and Communities}

Whether in networks or another form, Twitter data yield insight through a multitude of different analytical techniques. One such technique examines tweets as they flow through the network in “cascades.” Cascades follow a single tweet that is retweeted or similar tweets as they move across a network. The Twitter platform makes these analyses difficult, however, as retweets are connected to the original tweet, not the tweet that triggered the retweet \cite{ruths2014}. Researchers address this pitfall by using temporal sequencing to order and connect tweets or retweets. To achieve meaningful results, studies must sufficiently filter the tweets to establish that sequential tweets are related in content as well as time, which undermines representativeness, as discussed above \cite{alvarez2015,borge2015,weng2013}.

Another common technique examines tweet content. Alvarez et al. \cite{alvarez2015} analyze their data for its social and sentiment content using semantic and sentiment analytic algorithms that analyzes tweets based on a test set. The authors use this technique to draw conclusions of individual users’ opinions of the 15M movement in Spain by analyzing up to 200 authored tweets on the topic per user. This technique holds great promise for future studies of political activity, and indeed any activity, on Twitter. \cite{borge2015} uses a less sophisticated solution towards a similar goal: they characterize users as either for or against military intervention in Egypt. The authors attempt to show changes in opinion, and so cannot not rely on comprehensive opinion from a mass of past tweets as done in \cite{alvarez2015}. Instead, \cite{borge2015} uses coded hashtags to indicate users' opinions. Although this technique allows for discernable changes in opinion, the authors establish a dichotomy that threatens to oversimplify users' opinions.

Community detection is another key analytical tool for Twitter researchers. Using network topology or node (user or tweet) content, researchers can cluster similar nodes and provide insight into social systems on a macroscopic scale. There are a variety of techniques, each with its own set of strengths and weaknesses. \cite{weng2013} uses the Infomap algorithm \cite{rosvall2008} and test the robustness of their results by applying a second community detection technique, Link Clustering. \cite{conover2012} uses a combination of two techniques, Rhaghavan's label propagation method \cite{raghavan2007} seeded with node labels from Newman's leading eigenvector modularity maximization \cite{newman2006}. The authors selected this combination of methods because it ``neatly divides the population ... into two distinct communities.'' Yet, the authors fail to defend these observations rigorously in their paper. \cite{beguerisse2014}, on the other hand, effectively defends their decisions in setting resolution parameters for the Markov Stability method \cite{schaub2012}. The authors also use community detection creatively in conjunction with a functional role-determining algorithm to assign ``roles'' to users without a priori assignments of those groups. \cite{borge2015} selects an apt community detection method that corresponds well with their objectives: to follow changes in polarity over time, the authors use label propagation, whereby nodes spread their assigned polarity. This method allows for seeding with nodes of known belief---useful in monitoring the progression of the Egyptian protests on Twitter, as many important actors' positions were publicly known. Yet this decision too comes with a cost: the authors program the label propagation to allow for only two polarities: Secularist or Islamist, even though they acknowledge that a third camp likely existed, namely supporters of deposed Hosni Mubarak.

While community detection is still considered as an open question in network science, both at the definition and algorithmic implementation levels \cite{fortunato2010}, many papers use one or more of these methods without enough care to make sure that the methods and definitions that they are using in their specific problem is well justified. 

\subsection{Experiments}

Twitter also lends itself as an experimental platform for researchers to implement controlled studies of social phenomena. In particular, \cite{garcia2014} and \cite{weng2013} seek to predict viral memes on Twitter using network topology and activation in linked users and communities, respectively. \cite{coppock2015} runs two experiments on inducing political behavior on Twitter using different types and phrasing of messages. In all cases, authors necessarily use controls in their experimental context. \cite{garcia2014} creates a null distribution of tweets with randomly shuffled timestamps to distinguish the effect of user centrality from user tweeting rate. \cite{weng2013} uses two baseline models to quantify the predictive power of their community-based model. The authors use a random guess and community-blind predictor, against both of which the model is highly statistically significant. \cite{coppock2015}, with a true experimental design, offer an extensive discussion of experimental controls on Twitter. The platform has inherent limitations for public tweet experiments because there is no effective way to separate experimental and control users given an inherently interconnected network structure. But the authors design their study to use direct messages to selected users as the experimental variable. The authors even tweaked and repeated the study to improve randomization in the control. Such a methodology makes \cite{coppock2015} an example of a particularly strong experimental Twitter paper.

\subsection{Conjecture}
As we have seen, Twitter provides researchers myriad analytical techniques. Methodological choices as to which techniques to use present a fundamental challenge for researchers. They must select and properly defend their choice of methods that both work and fit their theoretical objectives. As we have noted above, there are numerous instances where researchers will do better jobs than others are achieving a methodological fit and defending it in their studies. Some researchers may face the temptation to extend analyses to produce exciting results, but do so at the expense of sound methodologies. Future Twitter research would be well served to stress defensible, rigorous methodologies that are couched within existing theoretical literature from the social sciences, something that is rare today.

\section{What did we learn?}

Taken collectively, the reviewed papers offer considerable insight into political activities conducted on the Twitter platform, through analyses that examine political action in the abstract and others that offer case studies of concrete political action. These insights particularly address the roles of communities and individual users, connections between such entities, as well as the content they tweet. Predictive models take these insights and offer tools for, perhaps, understanding political action in real-time. \cite{garcia2014} uses a sensor group of central users to predict virality of content, and extend this predictive sensor beyond Twitter to Google searches. \cite{weng2013} uses connection topology to predict virality, although the predictive model is not extended to other content. \cite{gonzalez2011} observes viral tweets emerge from randomly distributed seed users, indicating exogenous factors determine the origins of viral content. Taken together, these three studies offer an understanding of mass communication on Twitter: viral content tends to originate randomly across the platform, reach more central users first, and spread across communities more easily than non-viral content. Theoretical explanations of what makes viral content in the first place, however, is lacking in these analyses, and warrants further attention.

Topology also offers insights into its embedded users. \cite{beguerisse2014} uses topographical analyses to reveal flow based roles, interest communities, and individual vantage points without a priori assignment. \cite{conover2012} assigns political leaning and then examine differences in partisan topologies in communities, tweeting activity, retweeting behavior, and mentions. Both approaches offer insight into political behavior using topology, with different strengths. The techniques used in \cite{beguerisse2014} are quite useful when the partisan landscape on a particular issue is unknown; The approach in \cite{conover2012}yields greater understanding of known divisions. 

Topology is not the sole determinant of activity, however, and tweet content analyses offer a second means of understanding political activity on Twitter. \cite{alvarez2015} finds that, in the context of the Spanish 15M indignados, tweets with high social and negative content spread in larger cascades. Tweet content also readily lends itself to analyses which link Twitter with offline phenomena. \cite{borge2015} and \cite{gonzalez2011} find that, in 2013 Egyptian protests and Spanish 15M protests, respectively, real world events impact tweeting behavior. \cite{coppock2015} successfully induces off-Twitter behavior using the content of tweets. Content analyses offer insight into non-platform-dependent political activity. 

Topology and content are distinct analyses. Research that combines the two to answer a single question can yield robust results. Several papers attempt this, Borge-Holthoefer et al. \cite{borge2015} most successfully. The authors use content analysis to classify tweets and users into opinion groups, and then create temporally based retweet networks to follow changes in the activity and composition of those opinion groups. Alvarez et al. \cite{alvarez2015} use content analysis of observed network topological phenomena, e.g., cascades, to quantify the social and emotional effects of content on sharing outcomes. Beguerisse-Dias et al. \cite{beguerisse2014} too combine methodologies, although less rigorously: they use word clouds to label topologically derived network communities. 

In this vein, many of the above mentioned papers could benefit from incorporating mixed methodologies and drawing on each others’ analyses. Future research papers should seek to emulate the approach in \cite{borge2015}. Further use of sentiment analyses from \cite{alvarez2015} would render even more robust results. Additional joint content and topology analyses would be even more useful: would using \cite{garcia2014}'s central users in communities, i.e., incorporate Weng et al.’s methods \cite{weng2013}, result in to more precise virality predictor? Would adding content analysis as used in \cite{alvarez2015} further improve precision? If holistic understanding of social phenomena is researchers’ goal, future efforts should seek to incorporate not one but numerous methodologies in pursuit of that end.

\section{Conclusion}
The papers considered in this minireview offer several important considerations on the state of Twitter research into social phenomena. What was once the arena of solely political scientists and sociologists, political action and social phenomena have now become research topics for computer scientists and social physicists. New disciplines have much to offer social research, as indicated in the methodology review of our sample papers; yet, these methodologies are often divorced from underlying social theory. Thus far, Twitter studies offer primarily observational ---not explanatory--- analyses. Greater dialog between theory and methods, as well as a holistic use of all available methodologies, is needed for data science to truly offer insight into our social world, both on Twitter and off it.

\section*{Disclosure/Conflict-of-Interest Statement}
%Frontiers follows the recommendations by the International Committee of Medical Journal Editors (http://www.icmje.org/ethical_4conflicts.html) which require that all financial, commercial or other relationships that might be perceived by the academic community as representing a potential conflict of interest must be disclosed. If no such relationship exists, authors will be asked to declare that the research was conducted in the absence of any commercial or financial relationships that could be construed as a potential conflict of interest. When disclosing the potential conflict of interest, the authors need to address the following points:
%•	Did you or your institution at any time receive payment or services from a third party for any aspect of the submitted work?
%•	Please declare financial relationships with entities that could be perceived to influence, or that give the appearance of potentially influencing, what you wrote in the submitted work.
%•	Please declare patents and copyrights, whether pending, issued, licensed and/or receiving royalties relevant to the work.
%•	Please state other relationships or activities that readers could perceive to have influenced, or that give the appearance of potentially influencing, what you wrote in the submitted work.
The authors declare that the research was conducted in the absence of any commercial or financial relationships that could be construed as a potential conflict of interest.

\section*{Author Contributions}
%When determining authorship the following criteria should be observed:
%•	Substantial contributions to the conception or design of the work; or the acquisition, analysis, or interpretation of data for the work; AND
%•	Drafting the work or revising it critically for important intellectual content; AND
%•	Final approval of the version to be published ; AND
%•	Agreement to be accountable for all aspects of the work in ensuring that questions related to the accuracy or integrity of any part of the work are appropriately investigated and resolved.
%Contributors who meet fewer than all 4 of the above criteria for authorship should not be listed as authors, but they should be acknowledged. (http://www.icmje.org/roles_a.html)
All authors listed, have made substantial, direct and intellectual contribution to the work, and approved it for publication.

\section*{Acknowledgments}

For providing useful feedback on the original manuscript we thank xxx xxx xxx.

%\section*{Supplemental Data}

%\bibliographystyle{frontiersinSCNS_ENG_HUMS} % for Science, Engineering and Humanities and Social Sciences articles, for Humanities and Social Sciences articles please include page numbers in the in-text citations
%\bibliographystyle{frontiersinHLTH&FPHY} % for Health and Physics articles
%\bibliography{twitter}

%%% Upload the *bib file along with the *tex file and PDF on submission if the bibliography is not in the main *tex file

%%% Use this if adding the figures directly in the mansucript, if so, please remember to also upload the files when submitting your article
%%% There is no need for adding the file termination, as long as you indicate where the file is saved. In the examples below the files (logo1.jpg and logo2.eps) are in the Frontiers LaTeX folder
%%% If using *.tif files convert them to .jpg or .png

%\begin{figure}
%\begin{center}
%\includegraphics[width=10cm]{logo2}% This is an *.eps file
%\end{center}
%\textbf{\refstepcounter{figure}\label{fig:02} Figure \arabic{figure}.}{ Enter the caption for your figure here. Repeat as necessary for each of your figures }
%\end{figure}

%%% If you don't add the figures in the LaTeX files, please upload them when submitting the article.

%%% Frontiers will add the figures at the end of the provisional pdf automatically %%%

%%% The use of LaTeX coding to draw Diagrams/Figures/Structures should be avoided. They should be external callouts including graphics.

\end{document}